\documentclass[12pt]{iopart}

\usepackage{graphicx}
\expandafter\let\csname equation*\endcsname\relax
\expandafter\let\csname endequation*\endcsname\relax
\usepackage{amsmath}
\usepackage{soul}

\usepackage{caption}
\usepackage{subcaption}

\usepackage{amssymb}
\usepackage{bm}
\usepackage{xcolor}
\usepackage{hyperref}
\hypersetup{
     colorlinks   = true,
     citecolor    = blue
}
\usepackage{xcolor}

\newcommand{\gv}[1]{\ensuremath{\mbox{\boldmath$ #1 $}}} 
\newcommand{\guv}[1]{\ensuremath{\mbox{\boldmath$ \hat{#1} $}}} 
\newcommand{\revision}[1]{{#1}}

\begin{document}

\title[BOUT++ simulation of thermal quench]{Electromagnetic turbulence simulation of tokamak edge plasma dynamics and divertor heat load during thermal quench}
 
\author{Ben Zhu$^1$, Xue-qiao Xu$^1$ and Xian-Zhu Tang$^2$}
\address{$^1$Lawrence Livermore National Laboratory, Livermore, California 94550, USA}
\address{$^2$Theoretical Division, Los Alamos National Laboratory, Los Alamos, New Mexico 87545 USA}
\ead{zhu12@llnl.gov}

\vspace{10pt}
\begin{indented}
\item[]June 2023
\end{indented}

\begin{abstract}
The edge plasma turbulence and transport dynamics, as well as the
divertor power loads during the thermal quench phase of tokamak
disruptions are numerically investigated with BOUT++'s flux-driven,
six-field electromagnetic turbulence model. Here transient yet
intense particle and energy sources are applied at the pedestal top to
mimic the plasma power drive at the edge induced by a core thermal collapse,
which flattens core temperature profile.  Interesting features such
as surging of divertor heat load (up to 50 times), and broadening of
heat flux width (up to 4 times) on the outer divertor target plate,
are observed in the simulation, in qualitative agreement with
experimental observations.  The dramatic changes of divertor heat load
and width are due to the enhanced plasma turbulence activities inside
the separatrix.  Two cross-field transport mechanisms, namely the
$E\times B$ turbulent convection and the stochastic parallel
advection/conduction, are identified to play important roles in this
process.  Firstly, elevated edge pressure
gradient drives instabilities and subsequent turbulence in the entire
pedestal region.  The enhanced turbulence not only transports
particles and energy radially across the separatrix via $E\times B$
convection which causes the initial divertor heat load burst, but also
induces an amplified magnetic fluctuation $\tilde{B}$.  Once the
magnetic fluctuation is large enough to break the magnetic flux
surface, magnetic flutter effect provides an additional radial
transport channel.  In the late stage of our simulation,
$|\tilde{B}_r/B_0|$ reaches to $10^{-4}$ level that completely breaks
magnetic flux surfaces such that stochastic field-lines are directly
connecting pedestal top plasma to the divertor target plates or first
wall, further contributing to the divertor heat flux width broadening.

\end{abstract}

%
%
%
\maketitle
%
%

\section{Introduction}

Tokamak disruptions are a major concern for future reactors such as
ITER~\cite{sugihara2007disruption,lehnen2015disruptions} as the rapid
release of plasma thermal and magnetic energy threatens machine
integrity. A tokamak disruption is known to occur in two phases, a
thermal quench phase characterized by the rapid loss of stored thermal
energy in the core and a subsequent slower current quench phase
characterized by the loss of stored magnetic energy (i.e., decay of
plasma current). It is commonly expected that a naturally occurring
major disruption on ITER would produce a core thermal collapse in the
order of a millisecond~\cite{riccardo2005timescale,hender2007mhd,lehnen2013disruption}.
Since an ITER plasma has roughly 200-300 mega-joule (MJ) of thermal energy, this would
correspond to 200-300 gigawatts (GW) of plasma power exhaust if all of
this energy is immediately deposited onto the divertor and first wall.
This can be compared with a steady-state ITER burning plasma that
would have around 150 megawatts (MW) of plasma power load on the
plasma-facing components (PFC). In a mitigated thermal quench, on the
condition that a sufficiently long lead time on disruption precursors
allows the prompt injection of high-Z pellets into the core plasma,
most of this plasma power is to be radiatively exhausted to the PFC
with good uniformity.  In case that option is not available due to the
short lead time or that the impurity radiative exhaust does not perform as
desired, a significant fraction of the plasma power would arrive at
the PFC in the form of plasma kinetic energy. This is the scenario we
will investigate in this paper.

Experiments on current tokamaks reveal that the plasma energy released
from the core temperature collapse, which defines the duration of the
thermal quench, does not arrive at the divertor and first wall
immediately. Instead, there can be a sizable delay between the core
temperature collapse and the heat pulse measured on the divertor
plate.  What likely happened is that the original edge/boundary
plasma, which is relatively cold compared with core plasma before
a disruption, serves as a buffer that temporally stores the thermal
energy released from the core.  The effectiveness of the buffer can be
aided by the massive gas release from the divertor/wall that have
previously stored significant amount of irradiating plasma ions in a
short pulse machine, or from the opening of the safety gas valves when
they are tripped by the first indication of excessive divertor
heating. The resulting higher edge/boundary plasma density increases
its heat capacity to hold the plasma energy released from the core
thermal collapse.  This physical scenario is consistent with the
experimental observation that the divertor heat load rise time
$\tau_r$ often matches the thermal quench duration $\tau$; while the
energy deposition time (i.e., the decay time of divertor heat load
after it reaches the maximum) is longer than $\tau.$~\cite{hender2007mhd}

Another interesting and important experimental observation on existing
tokamaks, is that the divertor heat load could increase tenfold or
more~\cite{hender2007mhd,lehnen2013disruption}, but the divertor heat
flux width $\lambda_q$ could broaden a few
times~\cite{lehnen2015disruptions}. The most intuitively obvious
causes are (i) field line stochasticization that smears and broadens
the plasma wetting area on the divertors and (ii) the enhanced
turbulent transport that can increase the scrape-off layer width and
thus divertor plasma wetting area. For the former, it is known that 3D
magnetic perturbation can produce complicated strike point pattern on
the divertor plates, both poloidally and
toroidally.~\cite{evans2002modeling,evans2005experimental,evans2007experimental}
For the latter, it was previously
understood that for large devices like ITER, the scrape-off layer width
would be set by turbulent as opposed to neoclassical
transport in steady-state operation.~\cite{chang2017gyrokinetic,li2019prediction,wang2021fluid,he2022prediction}
In the context of tokamak disruptions,
naturally occurring disruptions are almost always associated with
significant 3D MHD activities that break the magnetic flux surfaces,
so the mechanism of (i) is to be expected for divertor heat flux
broadening.  A distinction can be made here depending on whether the
disruption is initially driven by internal or external MHD modes. If
the disruption is dominated by external modes, the same
thermal-quench-inducing MHD activities would also set the divertor
heat flux width broadening. For a core thermal collapse induced by
internal MHD modes, there is the possibility that the first phase of
the core thermal collapse is a flattening of the core temperature
profile, which results in significant steepening of the edge pressure
profile.  This paper focuses on this second scenario and investigates
how the pedestal and scrape-off layer respond to an overloading of
plasma heating power from upstream core plasmas. The physics aims are to
understand how the excessive power loading at the edge can drive
divertor heat flux broadening in both space and time.

Our work can be contrasted with previous thermal quench simulations
using extended MHD codes such as
NIMROD~\cite{izzo2008magnetohydrodynamic}, M3D-C1~\cite{ferraro20183d}
and JOREK~\cite{nardon2021thermal}, which captures the global MHD
activities but not the edge turbulence commonly observed in pedestal
and scrape-off layer transport studies.  The approach chosen here is
also different from the gyrokinetic and fully kinetic simulations of
tokamak thermal quench. \revision{For examples, the collisionless plasma transport process in a prescribed stochastic 3D circular field was studied by the electrostatic gyrokinetic code GTS~\cite{yoo2021collisionless}, and a fully kinetic 1D3V calculation was performed to explore the dynamics of electron and ion thermal collapse in an open field-line by VPIC~\cite{zhang2023cooling}. In this study, thermal quench in a diverted edge plasma is investigated with the BOUT++ global six-field turbulence model~\cite{zhu2021drift} in a realistic tokamak geometry.} 
This kind of simulation resolves the electromagnetic edge turbulence, and
allows the quantification of the distinct roles and comparative
importance of turbulence-induced $E\times B$ transport and the parallel
transport along stochastic field lines in setting the spatial and
temporal profile of divertor heat flux during a thermal quench.

The paper is organized as follows. The physics model and simulation
set up are first introduced in
Section~\ref{sec:model}. Section~\ref{sec:qodt} presents the BOUT++
simulated divertor heat load evolution. The observed surging of heat
load and broadening of heat flux width are qualitatively consistent
with experimental observations. Section~\ref{sec:turb} studies the
role of edge turbulence inside the separatrix in setting the
downstream divertor heat load.  Section~\ref{sec:bpert} focuses on the
impact of turbulence induced magnetic fluctuation and how the
amplified $\tilde{B}_r$ changes the edge magnetic topology.
Section~\ref{sec:flux} further analyzes the contributions of $E\times
B$ turbulent convection and stochastic parallel advection and/or
conduction to radial particle and heat flux in this case.  Finally,
Section~\ref{sec:con} summarizes and discusses our key findings.

\section{Physics model and numerical setup}\label{sec:model}

\subsection{Physics model}

The numerical tool used in this paper is the six-field electromagnetic
turbulence model within BOUT++ framework~\cite{zhu2021drift} based on
the drift-reduced two-fluid Braginskii
equations~\cite{simakov2003drift}. 
This model has been extensively used to study ion-scale ($k\rho_i\ll 1$), low
frequency ($\omega\ll\omega_{ci}$) turbulence in tokamak edge plasmas;
and a similar set of equations is implemented in other 3D fluid-based
electromagnetic edge turbulence codes, such as
GBS~\cite{ricci2012simulation}, GDB~\cite{zhu2018gdb}, and
GRILLIX~\cite{stegmeir2018grillix}.  In this model, six independent
but nonlinearly coupled variables -- ion density $n_i$, electrostatic
potential $\phi$, ion parallel velocity $V_{\parallel,i}$, perturbed
parallel magnetic flux $A_\parallel$, electron and ion temperature
$T_{e,i}$, are evolved as
\begin{align}
\label{eq:density}
\frac{\partial}{\partial t}n_{i} &= -\left(\frac{1}{B}\boldsymbol{\hat{b}}\times\boldsymbol{\nabla}_{\perp}\phi+V_{\parallel i}\boldsymbol{\hat{b}}\right)\cdot\boldsymbol{\nabla} n_{i}
-\frac{2n_{i}}{B}\boldsymbol{\hat{b}}\times\boldsymbol{\kappa}\cdot\boldsymbol{\nabla}_{\perp}\phi+\frac{2}{ZeB}\boldsymbol{\hat{b}}\times\boldsymbol{\kappa}\cdot\boldsymbol{\nabla}_{\perp}P_{e} \nonumber \\
&-n_{i}B\nabla_{\parallel}\left(\frac{V_{\parallel i}}{B}\right)+\frac{B}{Ze}\nabla_\parallel\left(\frac{J_\parallel}{B}\right)+S_n,\\
\frac{\partial}{\partial t}\varpi & = -\left(\frac{1}{B}\boldsymbol{\hat{b}}\times\boldsymbol{\nabla}_{\perp}\phi+V_{\parallel i}\boldsymbol{\hat{b}}\right)\cdot\boldsymbol{\nabla}\varpi + B^{2}\boldsymbol{\nabla}_\parallel \left(\frac{J_{\parallel}}{B}\right)+2\boldsymbol{\hat{b}}\times\boldsymbol{\kappa}\cdot\boldsymbol{\nabla}P \nonumber \\ 
& -\frac{1}{2\Omega_{i}}\left[n_{i}Ze\boldsymbol{V_{D_i}}\cdot\boldsymbol{\nabla}\left(\nabla_{\perp}^{2}\phi\right)-m_{i}\Omega_{i}\boldsymbol{\hat{b}}\times\boldsymbol{\nabla}n_{i}\cdot\boldsymbol{\nabla}V_{E}^{2}\right] \nonumber \\
&+\frac{1}{2\Omega_{i}}\left[\boldsymbol{V_E}\cdot\boldsymbol{\nabla}\left(\nabla_{\perp}^{2}P_{i}\right)-\nabla_{\perp}^{2}\left(\boldsymbol{V_E}\cdot\boldsymbol{\nabla}P_{i}\right)\right], \\
\frac{\partial}{\partial t}V_{\parallel i} &= -\left(\frac{1}{B}\boldsymbol{\hat{b}}\times\boldsymbol{\nabla}_{\perp}\phi+V_{\parallel i}\boldsymbol{\hat{b}}\right)\cdot\boldsymbol{\nabla}V_{\parallel i}-\frac{1}{m_{i}n_{i}}\nabla_\parallel P
-\boldsymbol{V_{D_i}}\cdot\boldsymbol{\nabla}V_{\parallel i}-\frac{V_{\parallel i}S_n}{n_i},\\
\frac{\partial}{\partial t}A_\parallel &= -\nabla_{\parallel}\phi+\frac{\eta_{\parallel}}{\mu_{0}}\nabla_{\perp}^{2}A_\parallel+\frac{1}{en_{e}}\nabla_{\parallel}P_{e}+\frac{0.71k_{B}}{e}\nabla_{\parallel}T_{e},\\
\frac{\partial}{\partial t}T_{i} & = -\left(\frac{1}{B}\boldsymbol{\hat{b}}\times\boldsymbol{\nabla}_{\perp}\phi+V_{\parallel i}\boldsymbol{\hat{b}}\right)\cdot\boldsymbol{\nabla}T_{i} +\frac{2}{3n_{i}k_{B}}\nabla_{\parallel}q_{\parallel i} +\frac{2m_{e}}{m_{i}}\frac{Z}{\tau_{e}}\left(T_{e}-T_{i}\right) \nonumber \\
& -\frac{2}{3}T_{i}\left[\left(\frac{2}{B}\boldsymbol{\hat{b}}\times\boldsymbol{\kappa}\right)\cdot\left(\boldsymbol{\nabla}\phi+\frac{1}{Zen_{i}}\boldsymbol{\nabla}P_{i}+\frac{5}{2}\frac{k_{B}}{Ze}\boldsymbol{\nabla}T_{i}\right)+B\nabla_{\parallel}\left(\frac{V_{\parallel i}}{B}\right)\right] \nonumber \\
&-\frac{4}{3\Omega_{i}}T_{i}V_{\parallel i}\boldsymbol{\hat{b}}\times\boldsymbol{\kappa}\cdot\boldsymbol{\nabla}V_{\parallel i}+\frac{2 S^E_i}{3n_i}-\frac{T_i S_n}{n_i}, \\
\label{eq:ti}
\frac{\partial}{\partial t}T_{e} & = -\left(\frac{1}{B}\boldsymbol{\hat{b}}\times\boldsymbol{\nabla}_{\perp}\phi+V_{\parallel e}\boldsymbol{\hat{b}}\right)\cdot\boldsymbol{\nabla}T_{e} + \frac{2}{3n_{e}k_{B}}\nabla_{\parallel}q_{\parallel e} -\frac{2m_{e}}{m_{i}}\frac{1}{\tau_{e}}\left(T_{e}-T_{i}\right) \nonumber \\
& -\frac{2}{3}T_{e}\left[\left(\frac{2}{B}\boldsymbol{\hat{b}}\times\boldsymbol{\kappa}\right)\cdot\left(\boldsymbol{\nabla}\phi-\frac{1}{en_{e}}\boldsymbol{\nabla}P_{e}-\frac{5}{2}\frac{k_{B}}{e}\boldsymbol{\nabla}T_{e}\right)+B\nabla_{\parallel}\left(\frac{V_{\parallel e}}{B}\right)\right] \nonumber \\
& +0.71\frac{2T_{e}}{3en_{e}}B\nabla_{\parallel}\left(\frac{J_{\parallel}}{B}\right)+\frac{2}{3n_{e}k_{B}}\eta_{\parallel}J_{\parallel}^{2}+\frac{2 S^E_e}{3n_e}-\frac{T_e S_n}{n_e}.
\end{align}
Here the magnetic curvature $\gv{\kappa}$ is defined as $\gv{\kappa}=\guv{b}\cdot\gv{\nabla}\guv{b}\simeq\guv{b}_0\cdot\gv{\nabla}\guv{b}_0$ with unit vector $\guv{b}_0=\gv{B}_0/|B_0|$ denoting the direction of equilibrium magnetic field $\gv{B}_0$. This approximation implies that the perturbed magnetic field $\tilde{\boldsymbol{B}}\simeq \nabla A_\parallel \times \guv{b}_0$ shall remain small comparing to the equilibrium magnetic field $\boldsymbol{B}_0$, i.e., $\tilde{\boldsymbol{B}}\ll|\boldsymbol{B}_0|$.
The vorticity $\varpi$ takes the form
\begin{equation}\label{eq:vort}
    \varpi= \frac{n_i m_i}{B}\left(\nabla_\perp^2\phi +\frac{1}{n_i}\boldsymbol{\nabla}_\perp\phi\cdot\boldsymbol{\nabla_\perp} n_i+\frac{1}{Zen_i}\nabla_\perp^2P_i\right),
\end{equation}
and the total parallel current density $J_\parallel=J_{\parallel 0}+j_\parallel$ consists of both equilibrium current density $J_{\parallel 0}$ \revision{provided by kinetic equilibrium reconstruction} and perturbed current density $j_\parallel=-\nabla_\perp^2A_\parallel/\mu_0$.
All the transport coefficients in our model follow the original Braginskii transport model~\cite{braginskii1965transport} except for the parallel thermal conductivities $\kappa_{\parallel e,i}$ which are ``flux-limited" by
\begin{equation}\label{eq:qfl}
    \kappa_{\parallel e,i}^\text{eff}=\frac{\alpha_{e,i} n_{e,i}v_{\text{th}e,i}L_\parallel \kappa_{\parallel e,i}^\text{B}}{\alpha_{e,i} n_{e,i}v_{\text{th}e,i}L_\parallel+\kappa_{\parallel e,i}^\text{B}}
\end{equation}
to ensure the parallel heat flux $q_{\parallel e,i}$ is bounded by the local free-streaming condition.
In Equation~\ref{eq:qfl}, $v_{\text{th}e,i}$ are the local electrons and ions thermal speeds; and the parallel characteristic length is given by $L_\parallel=2\pi qR$ with $q$ the safety factor. In our simulation, flux limiting coefficients are set to be $\alpha=0.5$ for both electrons and ions.
The external plasma sources terms $S_n,S_i^E,S_e^E$ provide volumetric particle and heat flux source to replenish the plasma loss at the divertor targets in transport time-scale simulations; or in this study, they are used to over-drive a quiescent edge plasma system as discussed below.

\subsection{Simulation setup}

\begin{figure}[h]
  \centering
  \includegraphics[width=0.3\linewidth]{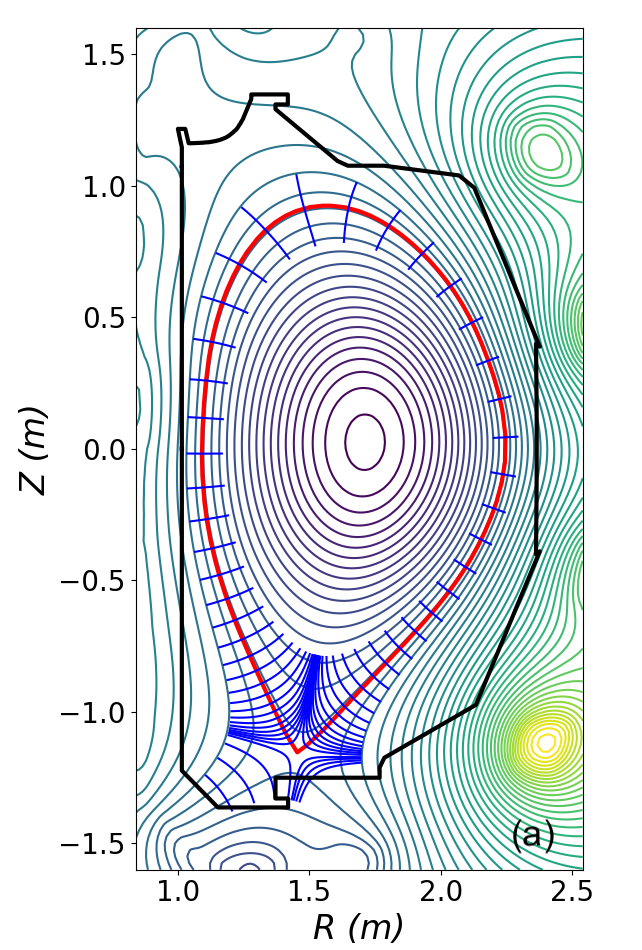}
  \includegraphics[width=0.6\linewidth]{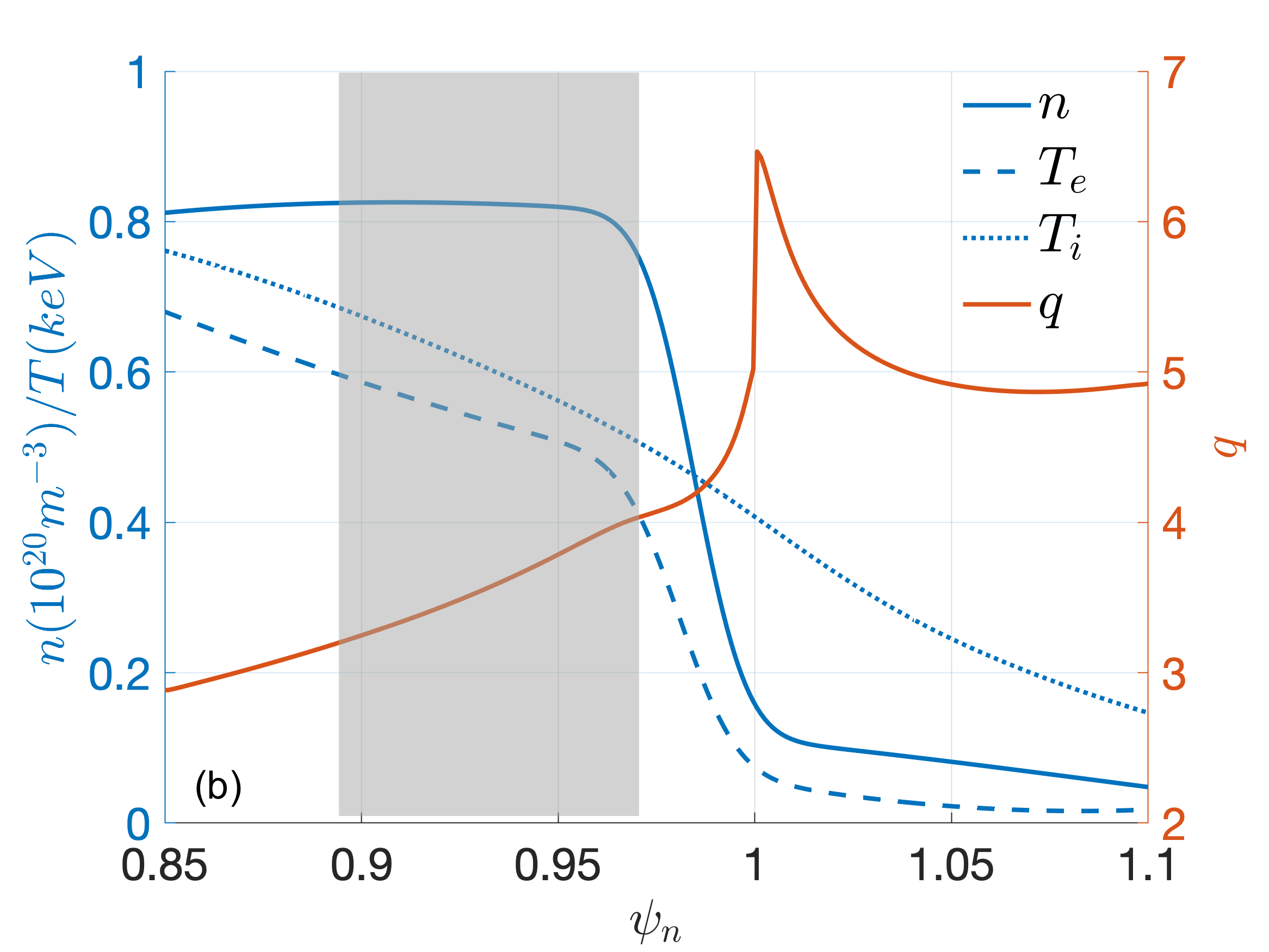}
    \caption{(a) DIII-D like lower single null discharge geometry and BOUT++ simulation domain (in blue lines), and (b) radial density and temperature profiles $n,T_{e,i}$ and safety factor $q$ of initial equilibrium with source region shaded.}%
   \label{fig:equilibrium}
\end{figure}

In this study BOUT++ simulation is started with a lower single-null H-mode plasma equilibrium with genetic DIII-D tokamak H-mode parameters which has been used for the validation of BOUT++ edge turbulence model on heat flux width scaling~\cite{xu2019simulations}. As illustrated in Figure~\ref{fig:equilibrium}, the radial simulation domain spans from $\psi_n=0.85$ to $\psi_n=1.1$ and covers a large portion of pedestal top, pedestal or steep gradient region, as well as the scrape-off-layer. This equilibrium has profound density and electron temperature pedestals at $\psi_N=0.98$, $q_{95}=3.8$, and electron and ion temperature around $500$ eV while the density about $0.8\times10^{20}$ m$^{-3}$ at $\psi_n=0.95$. The mesh resolution used in this study is $(n_x,n_y,n_z)=(260,64,64)$ where $(x,y,z)$ represents the radial, field-line and toroidal directions respectively. For computational efficiency only one-fifth of torus is simulated. We assume that plasma perturbation is zero at the outer radial boundary, i.e., the outer wall is ``far" away from the perturbed region such that $\tilde{n}_i=\tilde{T}_{e,i}=\tilde{\phi}=\tilde{V}_{\parallel i}=\tilde{j}_\parallel=\nabla_\perp^2A_\parallel=0$ at $\psi_n=1.1$; while for the inner radial boundary at $\psi_n=0.85$, homogeneous Neumann boundary condition is prescribed for plasma density, temperature and electrostatic potential, and the perturbed parallel current density, along with vanishing ion parallel velocity. In the field-line ($y$) direction, twist-shift periodicity is enforced inside the separatrix and Bohm sheath criterion is applied in the open field-line region~\cite{xia2015nonlinear}. The torodial ($z$) direction is naturally periodic. 

The initial edge equilibrium showed in Figure~\ref{fig:equilibrium} is marginally stable for peeling-ballooning modes. Therefore, a source-free simulation is first performed for $0.136$~ms to obtain a mild turbulent, non-disruptive edge plasma by allowing initial instabilities to grow and plasma profile to relax. In order to mimic extreme heat flux outflows from the core to the edge region when a core thermal quench occurs, volumetric flux source terms $S_n,S_i^E,S_e^E$ are turned on to simulate transient but intense flux arrived at the pedestal top. For simplicity, sources are assumed to be Gaussian shape radially and uniform along poloidal and toroidal directions,
\begin{equation}
    S_{n,E}=S_{n,E}^{0}\exp\left[-\frac{\left(\psi_n-\psi_0\right)^2}{2\delta_\psi^2}\right].
\end{equation}
The source centers at $\psi_0=0.934$ with a width $\delta_\psi=0.025$, so that nearly all the particles and power are deposited at the pedestal top.
In this simulation, the applied injection power is $1$ GW, equally partitioned between electrons and ions, and lasts for $85~\mu$s. Therefore, between $0.136$~ms to $0.221$~ms, a total of $85$ kJ energy, roughly $15\%$ of the total plasma thermal energy of a typical DIII-D H-mode plasma is injected. After $t=0.221$~ms, source terms are turned off once again.

\section{Divertor heat load evolution}\label{sec:qodt}

As mentioned in Section~\ref{sec:model}, there are three phases in our thermal quench simulation depending on whether the external plasma sourcing is applied or not. The initial $0.136$~ms source-free simulation is the first phase to set up a quiescent edge plasma system as the baseline; the second phase is from $0.136-0.221$~ms when the intensive heating and fueling are turned on; the third source free phase is for $t>0.221$~ms. 

\begin{figure}[h]
  \centering
  \includegraphics[width=0.45\linewidth]{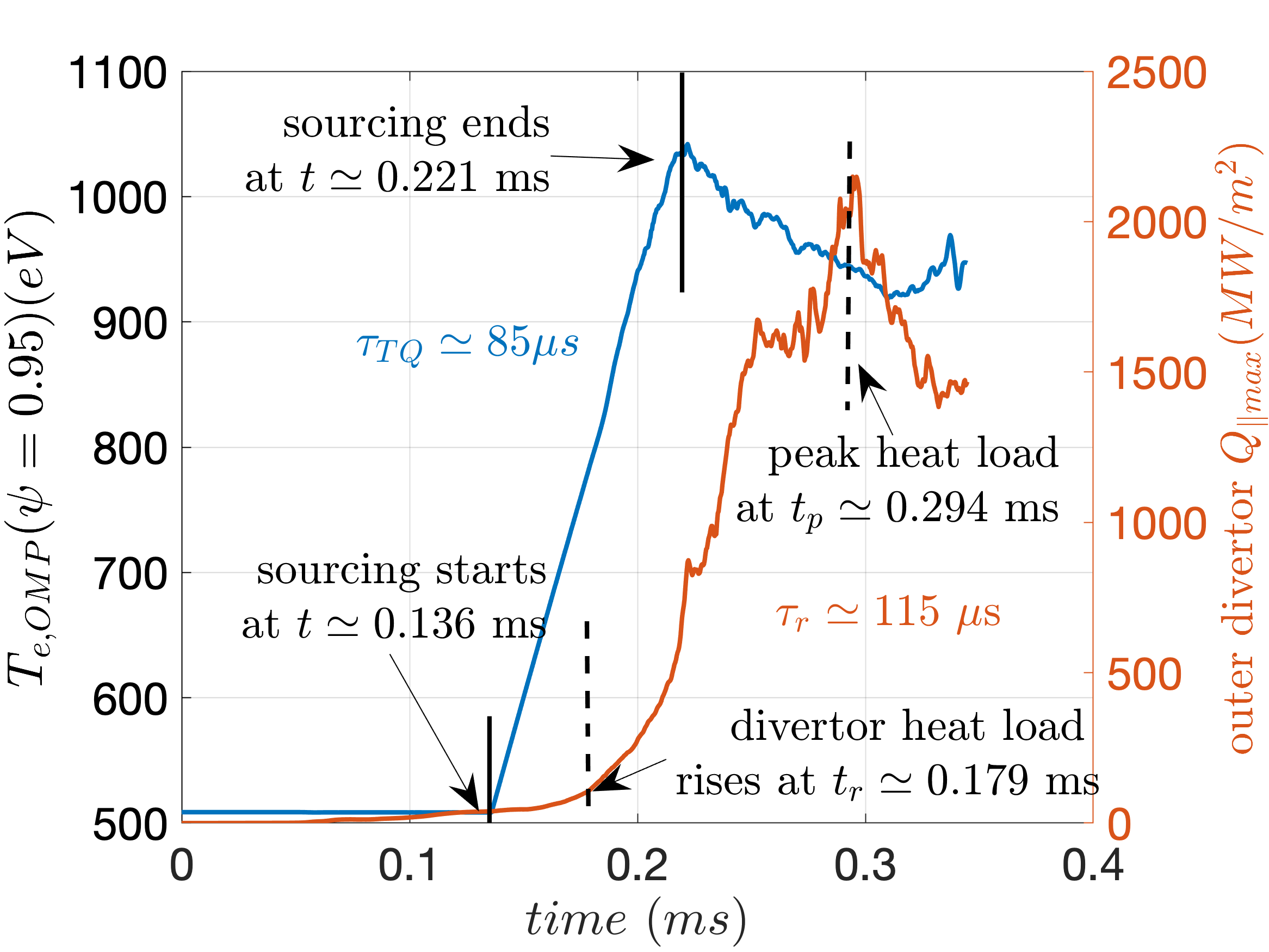}
  \includegraphics[width=0.45\linewidth]{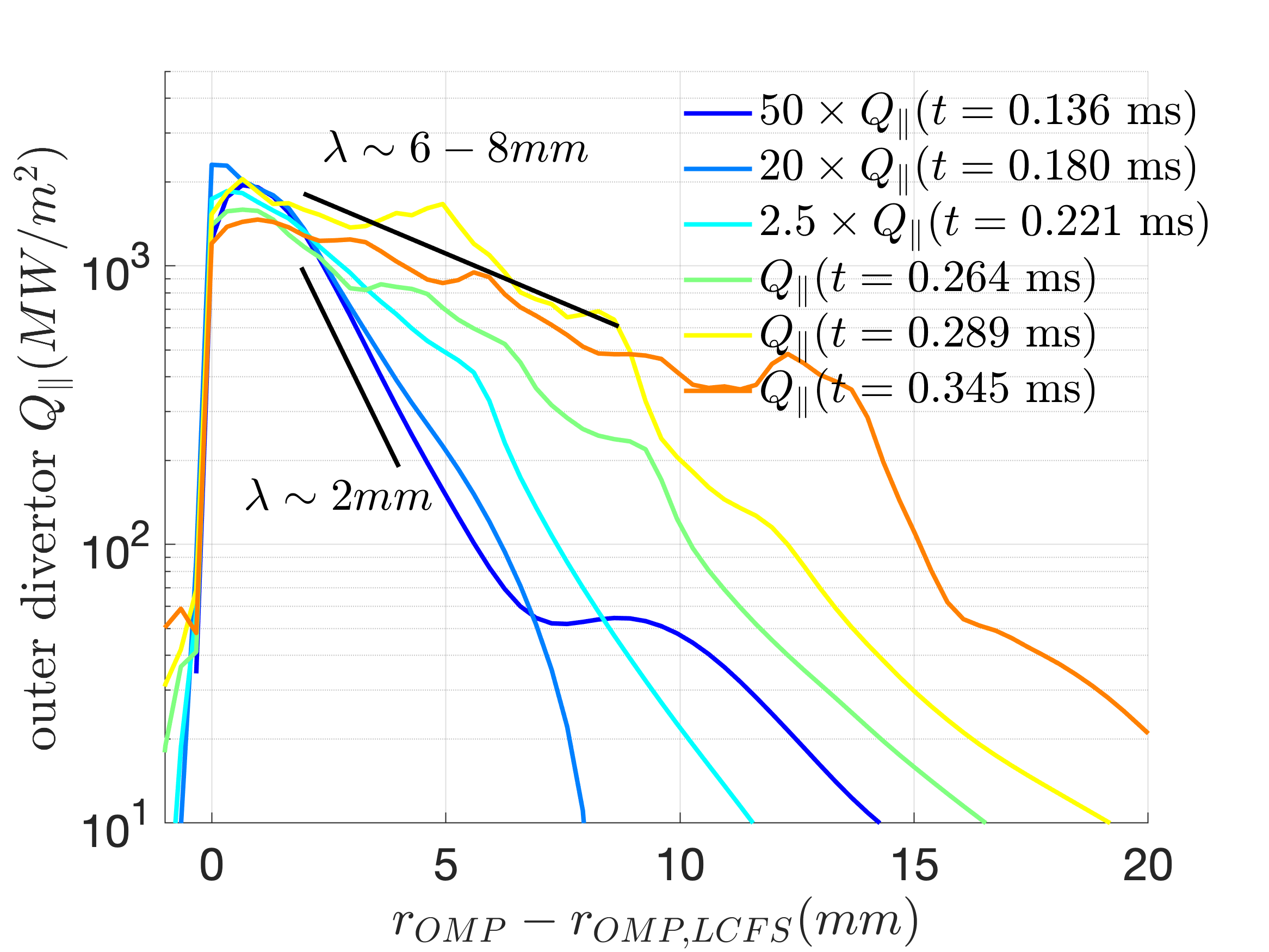}
    \caption{Toroidally averaged (left)outboard midplane pedestal top temperature evolution, and (right) outer divertor heat load profiles (mapped back to outboard midplane) during the simulation.}%
   \label{fig:T_omp_selected}
\end{figure}

We first examine the heat load on the divertor target plate in our simulation. 
Figure~\ref{fig:T_omp_selected}(a) shows the time history of electron temperature $T_{e,\text{OMP}}$ at the outboard mid-plane near the pedestal top ($\psi_n=0.95$), and the corresponding peak amplitude of toroidally averaged heat flux at outer divertor target plate. 
It can be seen that in phase 1 without external plasma sourcing, pedestal top electron temperature remains at $500$~eV level and the peak heat load slowly increases and statures at $40$~MW/m$^2$.
A monotonic, almost linear increase of pedestal top electron temperature is observed immediately after power injection is started. At $t=0.221$~ms, pedestal top electron temperature exceeds $1000$~eV, twice the initial value. When the power injection is turned off, pedestal top $T_{e,\text{OMP}}$ gradually decreases. 
The peak heat load at divertor target follows the evolution of pedestal top $T_{e,\text{OMP}}$ but with an apparent delay. Specifically, it starts to increase at $t_r\simeq0.179$~ms, (here we define the rising time begins at the 5\% of the maximum power load); and reaches its maximum value at $t_p\simeq 0.294$~ms. The rising time $\tau_r=t_p-t_r$ is roughly $115~\mu$s, which is slightly longer than $85~\mu$s — the thermal quench time $\tau_{TQ}$ we set in the simulation.

Figure~\ref{fig:T_omp_selected}(b) depicts the outer divertor heat load profiles at different times. Note here although the heat load is measured at divertor plate, it is projected back to the outer mid-plane to eliminate the influence of target angle with respect to the magnetic field. Without excessive sourcing, the heat flux width is about 2 mm and the peak amplitude is around 40 MW/m$^2$. These values are consistent with typical DIII-D H-mode discharges. Once the thermal quench occurs, heat flux amplitude increases and the width widens. As much as 50 times larger maximum heat load on the outer divertor target plate — from $40$~MW/m$^2$ to $2$~GW/m$^2$, and 3 to 4 times wider width — from $2$~mm to $6-8$~mm are observed in our simulation. Interestingly, the peak heat load drops after reaching its maximum value at $t=0.289$~ms; the heat flux width appears to remain at the similar level even at the later time (e.g., $t=0.345$~ms). 

We remark that divertor heat load features in our thermal quench simulation, such as $\tau_r\geq\tau_{TQ}$, peak heat load increasing by an order of magnitude and width broadening by  3-4 times, are in quantitative agreement with experimental observations on ASDEX-U~\cite{hender2007mhd} and TEXTOR~\cite{lehnen2015disruptions}.

\begin{figure}[h]
  \centering
  \includegraphics[width=\linewidth]{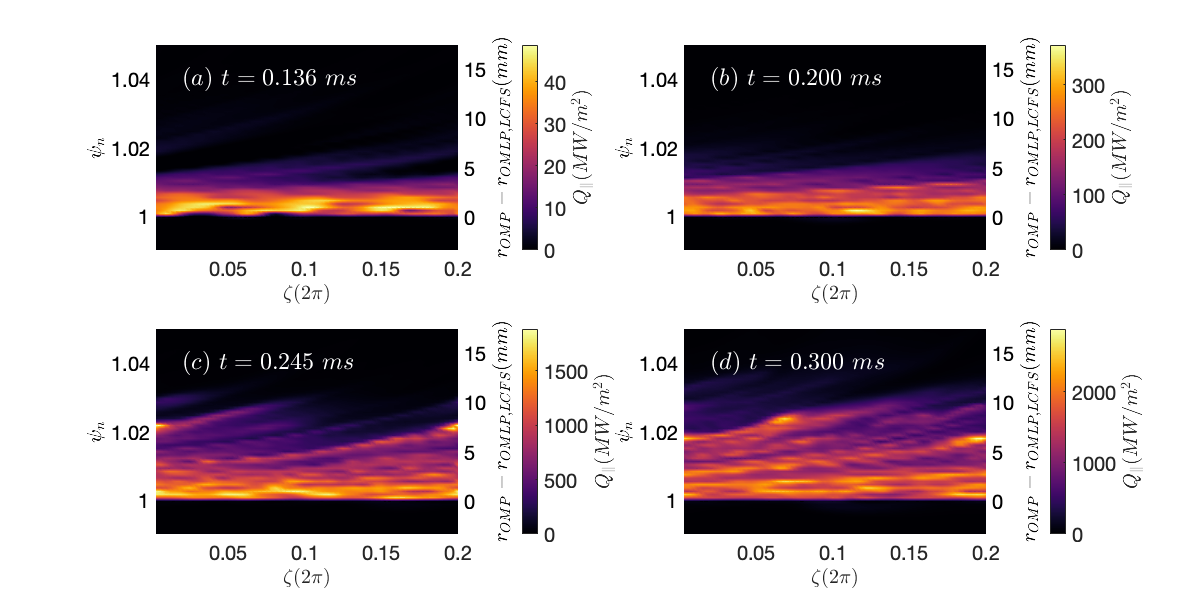}
  \caption{Outer divertor heat load profiles at selected times. Here $\zeta$ is the toroidal angle and $\psi_n$ is the normalized radial location.}%
  \label{fig:q2d}
\end{figure}

The non-smooth radial profiles of toroidally averaged $Q_\parallel$ at $t=0.264$~ms and afterwards suggest that $Q_\parallel$ could be nonuniform toroidally. For instance, multiple filaments may hit the divertor plate at different radial and toroidal locations, hence, causing multiple peaks in the toroidally averaged profile.
To better assess the toroidal distribution of $Q_\parallel$, full outer divertor target heat flux footprints at selected times are illustrated in Figure~\ref{fig:q2d} that highlight 2D structures of heat flux. Here heat flux are once again projected back to the outer mid-plane for a fair comparison.
At the early stage of the simulation, heat load is concentrated in the near scrape-off-layer (e.g., $1<\psi_n<1.01$) and only very weak striation patterns are observed. However, at $t=0.245$~ms and $t=0.3$~ms, helical striation patterns become clearly visible and the heat load extends towards the far scrape-off-layer region (e.g., $\psi_n>1.01$). These striation patterns on the divertor target indicate that the magnetic field structure may be altered in the simulation and hence impacts the divertor heat load~\cite{frerichs2010three}. Further discussion and analysis will be presented in Sections~\ref{sec:bpert} and \ref{sec:flux}.


\section{Enhanced edge turbulence fluctuation}\label{sec:turb}
Previous BOUT++ transport and turbulence studies find that turbulence activities inside the separatrix have strong influences on divertor heat exhaust~\cite{xu2019simulations}. In this study as we deliberately over-drive the edge plasma to mimic thermal quench process, we anticipate that turbulence would again play an important role in setting divertor heat load and width.

\begin{figure}[h]
  \centering
  \includegraphics[width=\linewidth]{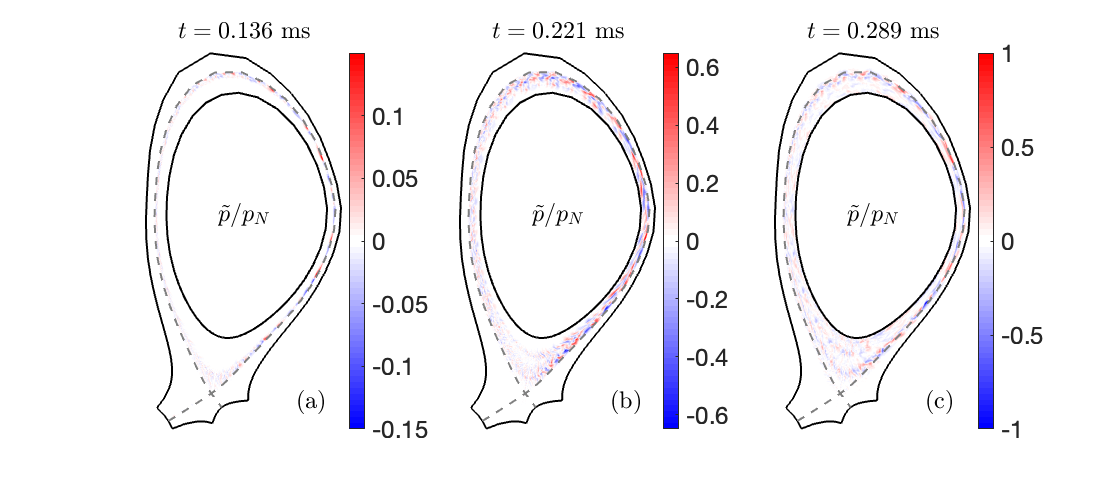}
    \caption{Poloidal snapshots of normalized pressure perturbation at (a) $t=0.136$~ms, (b) $t=0.221$~ms and (c) $t=0.289$~ms.}%
   \label{fig:ptilde}
\end{figure}

\begin{figure}[h]
  \centering
  \includegraphics[width=\linewidth]{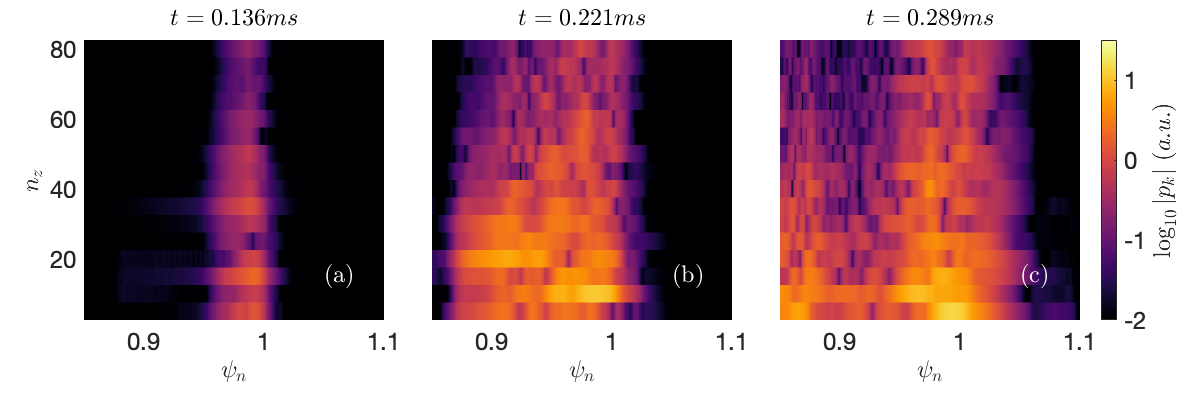}
    \caption{Toroidal mode number analysis of normalized pressure perturbation at the outboard mid-plane at (a) $t=0.136$~ms, (b) $t=0.221$~ms and (c) $t=0.289$~ms.}%
   \label{fig:pk}
\end{figure}

In the simulation, edge turbulence is found to enhance substantially with intensive heating and fueling as expected. The poloidal snapshots of normalized pressure perturbation show that the system is quiescent and the fluctuation is localized between the peak gradient region ($\psi_n\approx0.98$) and the separatrix prior to thermal quench (Figure~\ref{fig:ptilde}(a)). Once the plasma outflow from the core arrives at the pedestal top, it steepens the pressure profile which excites pressure-gradient-driven instabilities, resulting in not only stronger but also wider turbulent region across the entire pedestal region (Figure~\ref{fig:ptilde}(b)). Even after the outflow is turned off, the turbulence persists in the system and keeps spreading to the entire simulation domain. The overall fluctuation level increases about $8$ times comparing to the beginning (Figure~\ref{fig:ptilde}(c)). Note that the turbulence in general has the ballooning structure, i.e., stronger on the outboard side, we therefore analyze the pressure fluctuation spectrum at the outboard mid-plane for these three selected snapshots to understand whether the turbulence and transport characteristics are also changed as shown in Figure~\ref{fig:pk}. The system initially has a fairly low turbulent fluctuation level with the dominant mode $n_z=15$ at $\psi_n=0.99$. As the pedestal is elevated, at $t=0.221$~ms, there are two comparable amplitude modes, $n_z=10$ at $\psi_n=0.98$ and $n_z=20$ at $\psi_n=0.92$, corresponding to the two steep gradient regions caused by the Gaussian shaped source. At $t=0.289$~ms, the dominant mode has even lower $n_z=5$ at $\psi_n=0.99$. The shifting to lower $n_z$ modes implies that larger eddies are forming as the simulation progresses. These large eddies are more resilient to flow and magnetic shears; hence, they provide a more efficient radial transport channel to rapidly transfer particles and heat across the separatrix. Also, from Figure~\ref{fig:pk}, it is clear that turbulence spreading in our simulations is occurring not only in configuration space, but also in $k-$space.

\begin{figure}[h]
  \centering
  \includegraphics[width=0.8\linewidth]{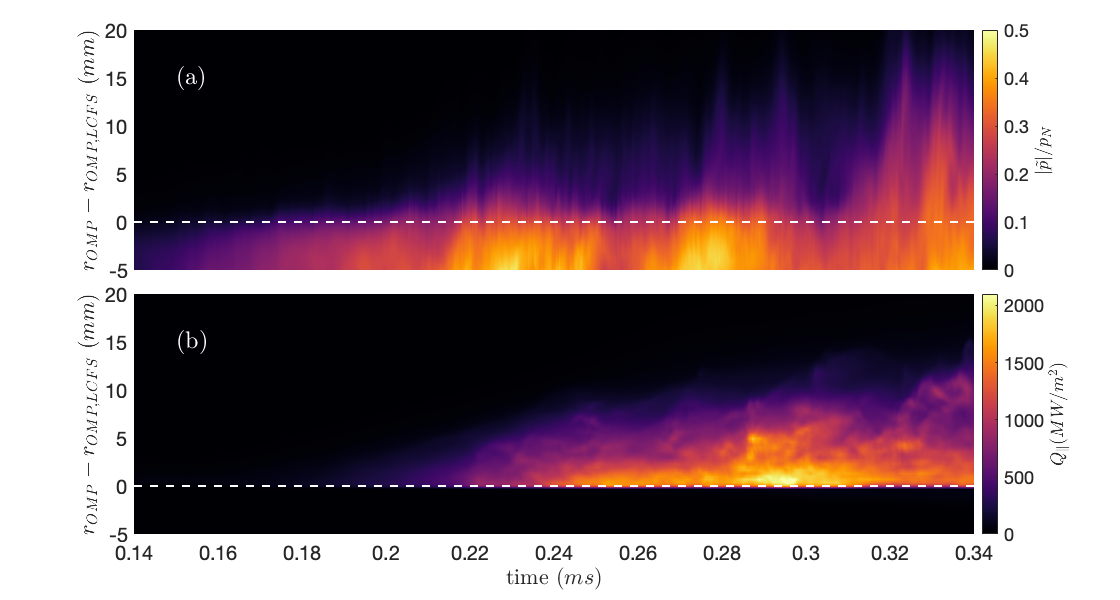}
    \caption{Normalized perturbed pressure at outboard mid-plane and the corresponding outer divertor heat load.}%
   \label{fig:pq_omp_evolution}
\end{figure}

\begin{figure}[h]
  \centering
  \includegraphics[width=0.5\linewidth]{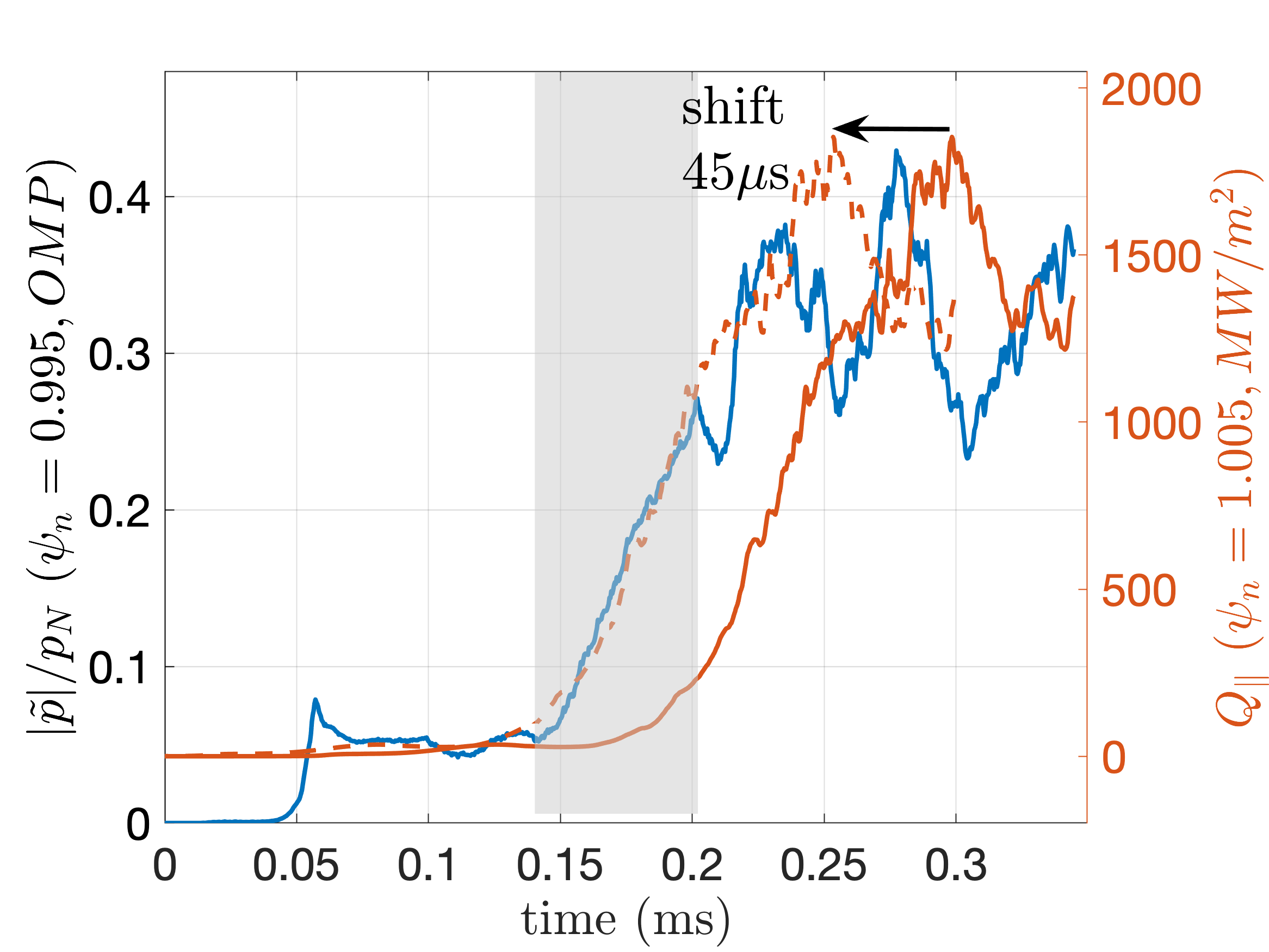}
    \caption{Evolution of outboard mid-plane total plasma pressure fluctuation inside the separatrix and the outer divertor target plate heat load.}%
   \label{fig:pq_corr}
\end{figure}

As we have confirmed that strong turbulence is present in the thermal quench simulation, we now look into the relation between the turbulence inside the separatrix and the heat load down at the divertor target. Figure~\ref{fig:pq_omp_evolution} displays the time evolution of radial normalized pressure perturbation profile at outboard mid-plane and the radial outer divertor heat flux profile, again mapped back to outboard mid-plane. Clearly divertor heat load response is lagging behind the mid-plane fluctuations. However, in general, there is a good correspondence between these two quantities -- as the outboard turbulence activity enhances, divertor heat load also increases.
To better illustrate the correlation between the outboard turbulence fluctuation and downstream divertor target heat flux, we plot both quantities at slightly different radial location in Figure~\ref{fig:pq_corr}. The pressure perturbation is just inside the separatrix at $\psi_n=0.995$ while the divertor heat load is just outside the separatrix at $\psi_n=1.005$. The vertical axis has been re-scaled to roughly match both quantities at the baseline level (i.e., $t\simeq 0.136$~ms) and their maximum (i.e., $t\simeq 0.294$~ms). At the early stage of divertor heat load rising time (e.g., $t=0.185-0.245$~ms), divertor heat load appears to well correlated with the mid-plane turbulence level after accounting for a roughly $45~\mu$s delay. 
This relatively short lag in time implies that the surging divertor heat load is largely influenced by electron parallel conduction. In our simulation, the (unperturbed) field-line length $L_0$ between outboard mid-plane and the outer divertor target just outside the separatrix is about 16 m and the local electron and ion temperature initially are about $150$ and $400$ eV respectively, so the sound speed and electron thermal speed are around $c_s=2\times10^5$ m/s and $v_{\text{th},e}=5\times10^6$ m/s. Based on these numbers, the characteristic parallel advection time is $t_\text{advection}=2L_0/c_s\simeq 160~\mu$s; while the electron free streaming (upper limit of electron thermal conduction) $t_{\text{fs},e}=L_0/v_{\text{th},e}\simeq 3~\mu$s. One could also estimate the characteristic parallel conduction time for electrons based on the effective (i.e., flux-limited) thermal conductivity in our simulation $\kappa_{\parallel, e}^\text{eff}\simeq 3\times10^6$~m$^2$/s, so $t_{\text{conduction},e}=L_0^2/\kappa_{\parallel,e}=85~\mu$s.
Therefore, parallel advection is too slow to explain the short lag between upstream (e.g., outboard mid-plane) turbulence enhancement and the downstream (e.g., divertor target) heat load rise in our simulation. Meanwhile, electron parallel thermal conduction process is fast enough to transport upstream power down to the target plate, and hence the primary contributor to the heat load surging. During this time period (i.e., $t=0.185-0.245$~ms at divertor target), heat flux width $\lambda_q$ increases from $2$~mm to about $4$~mm as illustrated in Figure~\ref{fig:T_omp_selected}(b). 
This over-driven thermal quench simulation is an example of heat flux width $\lambda_q$ scaling transitions from drift to turbulence dominant regime as the cross-field transport increases ~\cite{xu2019simulations}. In this turbulence dominant regime, classical heuristic drift-based model for scrape-off-layer scaling~\cite{goldston2011heuristic} becomes invalid due to the presence of large amplitude fluctuations along the field-line.

Another interesting observation from Figure~\ref{fig:pq_corr} is that after $t\simeq 0.2$~ms at upstream, or $t\simeq 0.245$~ms at downstream, the nice correspondence between outboard mid-plane turbulence and divertor heat load is less obvious despite the trend still matches. This is likely due to the complete destruction of magnetic flux surfaces which will be discussed in next Section.

\section{Amplified magnetic fluctuations}\label{sec:bpert}

In the electromagnetic drift-reduced Braginskii model, turbulence is almost always accompanied by magnetic fluctuations. Even for the electrostatic-instability-dominated turbulence, finite perturbed parallel vector potential $A_\parallel$ exists as long as there is a non-zero perturbed parallel current $j_\parallel$. Here we first revisit how the electromagnetic effect is incorporated in the drift-reduced Braginskii model. Without loss of generality, one can write the perturbed magnetic field $\gv{\tilde{B}}$ in term of perturbed vector potential $\gv{A}$
\begin{equation}
    \gv{\tilde{B}}=\nabla\times \gv{A}.
\end{equation}
In Coulomb gauge $\nabla\cdot \gv{A}=0$, so that $A_\parallel/L_\parallel \sim |\gv{A_\perp}|/L_\perp$, where $L_\perp,L_\parallel$ are the perpendicular and parallel characteristic lengths of the perturbation. If $L_\parallel\gg L_\perp$ for strongly magnetized plasmas, then $A_\parallel/|\gv{A_\perp}|\sim L_\parallel/L_\perp\gg 1$.
Similarly, the ratio between perpendicular and parallel component of the perturbed field is 
\begin{equation}
    \frac{|\gv{\tilde{B}}_\perp|}{\tilde{B}_\parallel}=\frac{|(\nabla\times \gv{A})_\perp|}{(\nabla \times \gv{A})_\parallel}\sim \cfrac{\frac{A_\parallel}{L_\perp}+\frac{|\gv{A}_\perp|}{L_\parallel}}{\frac{|\gv{A}_\perp|}{L_\perp}}\sim \cfrac{1+\frac{L_\perp^2}{L_\parallel^2}}{\frac{L_\perp}{L_\parallel}}\gg 1.
\end{equation}
Thus, drift-reduced Braginskii model neglects $\tilde{B}_\parallel$, and only keeps the $A_\parallel/L_\perp$ contribution to $\gv{\tilde{B}}_\perp$ such that the perturbed magnetic field is written as
\begin{equation} \label{eq:btilde}
    \gv{\tilde{B}}=\gv{\tilde{B}}_\perp=\nabla\times \left(A_\parallel \guv{b}_0\right).
\end{equation}
This expression may be further simplified as $\gv{\tilde{B}}\simeq \nabla A_\parallel \times \guv{b}_0$ if the characteristic length of equilibrium field $L_B\gg L_\perp$.
Therefore, strictly speaking, drift-reduced Braginskii model is ``semi" not ``full" electromagnetic, i.e., the background field is assumed to be incompressible ($\delta B_\parallel=0$). Nevertheless, the assumptions used in this derivation such as $L_\parallel \gg L_\perp$, $L_B\gg L_\perp$ are well satisfied for tokamak edge plasmas and also confirmed by extended full-MHD simulations~\cite{pamela2020extended}.

With the perturbed field expression, the parallel gradient operator in the model now has two separate terms,
\begin{equation}
    \nabla_\parallel f=\left(\guv{b}_0+\tilde{\gv{b}}\right)\cdot\nabla f=\guv{b}_0\cdot\nabla f-\frac{\guv{b}_0}{B}\times\nabla A_\parallel \cdot\nabla f.
\end{equation}
The first term on the right-hand-side denotes the parallel gradient of quantity $f$ along the unperturbed background field; while the second term represents the additional cross-field transport due to the perturbed field, often referred as ``magnetic flutter" effect. It is well known that magnetic flutter term could enhance radial transport level via direct contribution~\cite{xia2015nonlinear} or by indirect influencing of the turbulence saturation mechanism~\cite{rogers1997enhancement}. 
Here we will first look at the impact of magnetic flutter effect on the magnetic field structure and then quantify its influence on radial transport in next Section.

\begin{figure}[h]
  \centering
  \includegraphics[width=\linewidth]{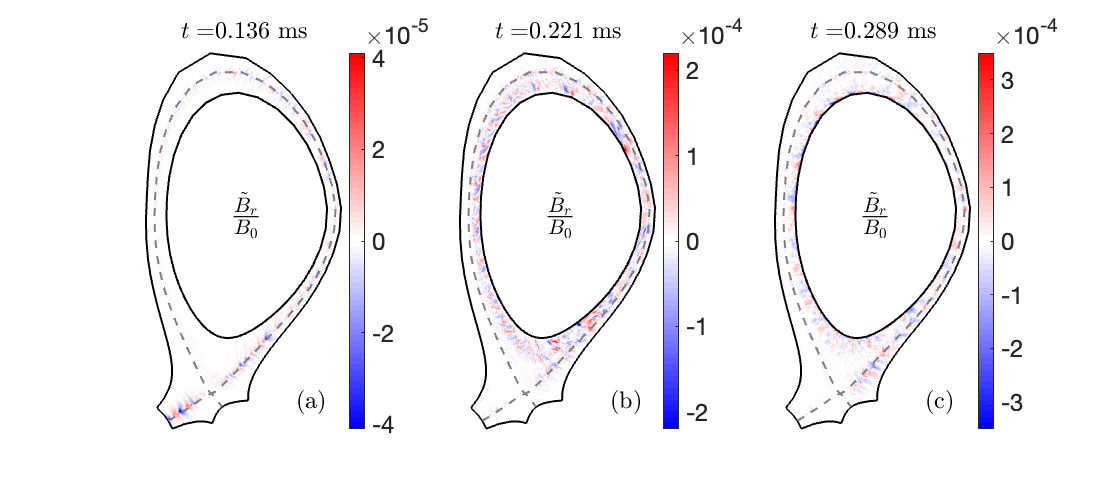}
    \caption{Poloidal snapshots of normalized radial component of perturbed magnetic field $\tilde{B}_r/B_0$ at (a) $t=0.136$~ms, (b) $t=0.221$~ms and (c) $t=0.289$~ms.}%
   \label{fig:brtilde}
\end{figure}

Figure~\ref{fig:brtilde} shows the poloidal snapshots of normalized radial component of perturbed magnetic field $\tilde{B}_r/B_0$ at $t=0.136$, $0.221$ and $0.289$~ms, same as the normalized perturbed pressure snapshots in Figure~\ref{fig:ptilde}. Although its value is often small, $|\tilde{B}_r/B_0|$ is an important metric of measuring the disturbance of local magnetic flux surface. Not surprisingly, $|\tilde{B}_r/B_0|$ increases significantly in the thermal quench simulation. Initially at $t=0.136$~ms without over-driving the system, $|\tilde{B}_r/B_0|\simeq 2\times 10^{-5}$ and the magnetic perturbation is primarily localized near the separatrix and concentrated on the outboard side, except for the inner divertor leg region. This result is consistent with the most BOUT++ edge turbulence simulations. However, at the end of intensive heating $t=0.221$~ms, the pedestal top region has a large magnetic perturbation with relative amplitude up to $2\times 10^{-4}$ uniformly distributed poloidally. Even after the sources were turned off, the perturbation level continues to grow and can reach to $10^{-3}$ level at the late stage of the simulation (e.g., $t>0.3$~ms).

\begin{figure}[h]
  \centering
  \includegraphics[width=\linewidth]{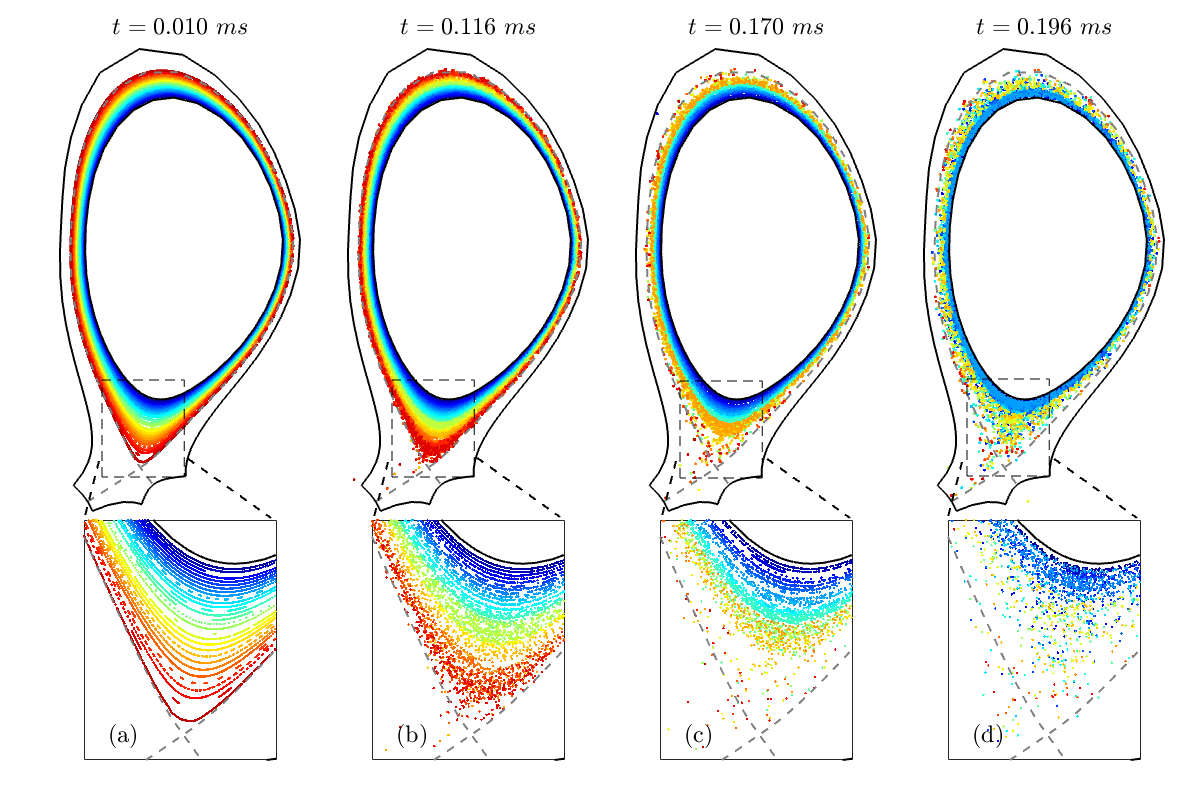}\\
    \caption{Evolution of Poincar\'e plots in BOUT++ thermal quench simulations. Snapshots are taken at (a) $t=0.010$~ms, (b) $t=0.116$~ms, (c) $t=0.170$~ms and (d) $t=0.196$~ms.}%
   \label{fig:poincare}
\end{figure}

The $O(10^{-4})$ level radial magnetic field perturbation has a drastic impact on the edge magnetic field topology as illustrated by the evolution of Poincar\'e plots in Figure~\ref{fig:poincare}. 
At the linear stage of the simulation (e.g., $t=0.01$~ms), perturbation induced by instabilities is negligible such that all the magnetic flux surfaces inside the separatrix are intact. When simulation enters fully nonlinear stage (e.g., $t=0.116$~ms), saturated turbulence disturbs the magnetic field-lines mostly near the separatrix and forms a layer of weakly stochastic field-lines; while the interior magnetic flux surfaces remain unbroken. However, shortly after excessive fueling and heating (e.g., $t=0.17$~ms), enhanced magnetic perturbation in the pedestal region distorts the local magnetic geometry. As a result, the stochastic layer originally localized near the separatrix starts to penetrate inward. At $t=0.196$~ms, almost all of flux surfaces in our simulation domain have been destroyed and the entire magnetic field within the ``closed flux region" becomes stochastic.

\begin{figure}[h]
  \centering
  \includegraphics[width=\linewidth]{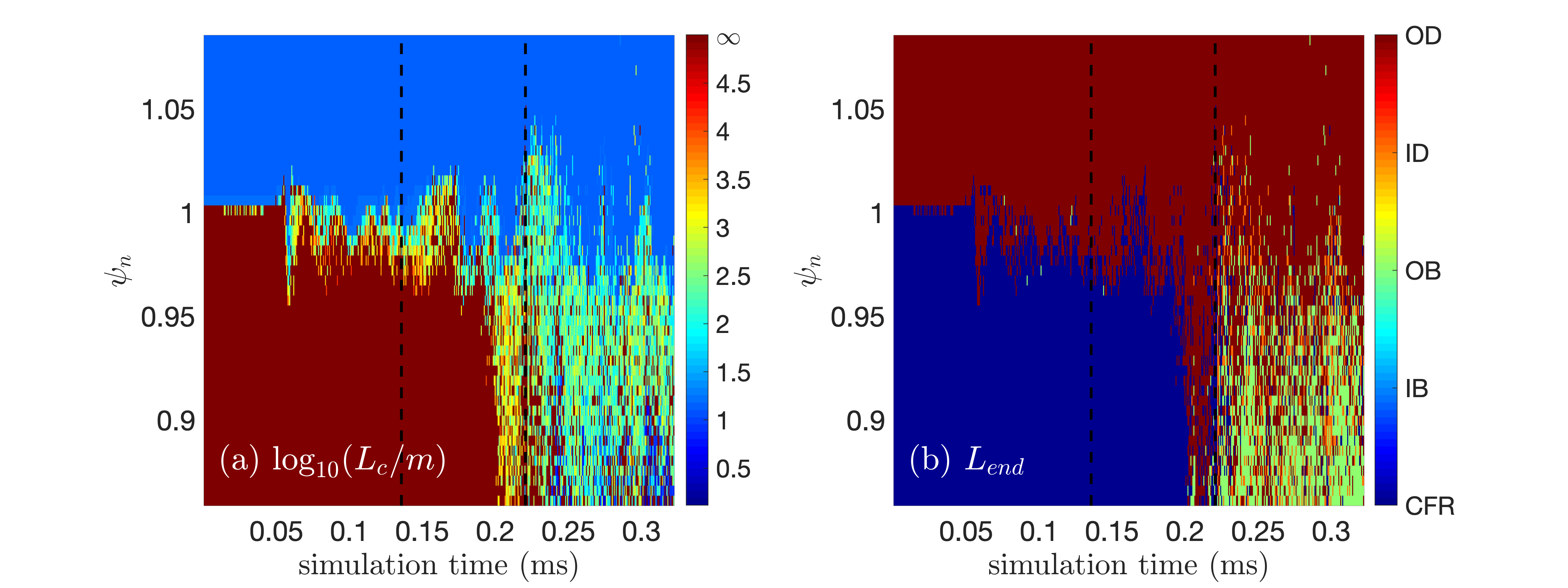}\\
    \caption{Evolution of (a) magnetic connection length $L_c$, and (b) final interception region of magnetic field-lines launched at outboard mid-plane. Here, CFR -- closed flux region, OB/IB -- outer/inner boundary, OD/ID -- outer/inner divertor.}%
   \label{fig:fieldline}
\end{figure}

The stochastization of magnetic field-lines in the tokamak edge region has a profound influence on the magnetic connection length $L_c$. Figure~\ref{fig:fieldline} shows the evolution of $L_c$ and final field-line interception region for magnetic fields located at outboard midplane in our simulation. In this analysis, we trace 48 radial uniformly distributed field-lines passing $\psi_n \in [0.859,1.086]$ on the outboard midplane at time $t$ in both $+y$ and $-y$ directions for a maximum of $250$ poloidal turns and record the field-line length $L_c^{(+)}$ and $L_c^{(-)}$ if it intercepts the divertor target or outer boundary. The connection length is defined as $L_c=\min(L_c^{(+)},L_c^{(-)})$; i.e., $L_c$ is the field-line length from the start point (i.e., outboard midplane in this case) to the nearest endpoint on a plasma facing component. If the field-line hits the inner boundary or remains in the closed flux region after $250$ poloidal turns, we consider $L_c^{(+)}$, or $L_c^{(-)}=\infty$. In an ideal equilibrium without any perturbation, 30 out of the 48, or, 62.50\% of the field-lines are started at $\psi_n<1$ and therefore should remain ``closed" with $L_c=\infty$; while the rest 37.50\% of the field-lines are considered as ``open" field-lines landing on the outer divertor and have $L_c\approx 20$~m. This scenario is approximately true till the onset of instability at around $t=0.05$~ms. During the nonlinear saturated stage (e.g., from $t=0.05$ to $0.136$~ms), the weakly stochastic layer near the separatrix reduces closed magnetic field-line percentage to 55.27\%; while the percentage of open field-lines ended at outer divertor target is 44.64\%, and the rest, which is a small portion (0.09\%), ended on the outer boundary. The typical magnetic connection length within this weakly stochastic layer is around $100-1000$ m. With the enhanced magnetic fluctuation in external sourcing phase ($t=0.136-0.221$~ms), it is clear to see the rapid expansion of stochastic layer in Figure~\ref{fig:fieldline}. At around $t=0.2$~ms, the entire pedestal region is stochastic. Overall, in this time period, percentage of field-lines remain closed is 43.12\% (with 0.5\% intercepting inner boundary); while 56.62\%, 0.1\%, and 0.17\% of the field-lines are ended on outer divertor, inner divertor and outer boundary respectively.
This stochastic magnetic field persists even after the heating and fueling are turned off. During $t=0.221-0.345$~ms, only 12.3\% of the field-lines are closed (4.99\% within the $0.85<\psi_n<1$ region and 7.30\% reach at inner boundary). The rest 87.7\% are open field-lines ending at outer divertor (64.86\%), inner divertor (8.18\%), and outer boundary (14.67\%). Interestingly, the majority of the field-lines that end at the outer boundary are from the inner region (e.g., $\psi_n<0.95$). This is because all the field-lines in this analysis are started from outboard mid-plane so that these near the separatrix (e.g., $\psi_n\simeq 0.98$) are likely entering the SOL immediately and hitting the outer divertor directly following the stochastic field-lines as the poloidal distance is limited and the magnetic perturbation level is relatively large (e.g., $|\tilde{B}_r/B_0|\sim 10^{-4}$). On the other hands, field-lines originated from the inner region has a large enough radial displacement to winding around, entering the SOL further away from divertor plates and hence ending at the outer boundary. The averaged magnetic connection length $L_c$ of open field-lines in $\psi_n<1.0$ region is on the order of $400$~m. 


\begin{figure}[h]
    \centering
    \includegraphics[width=0.45\linewidth]{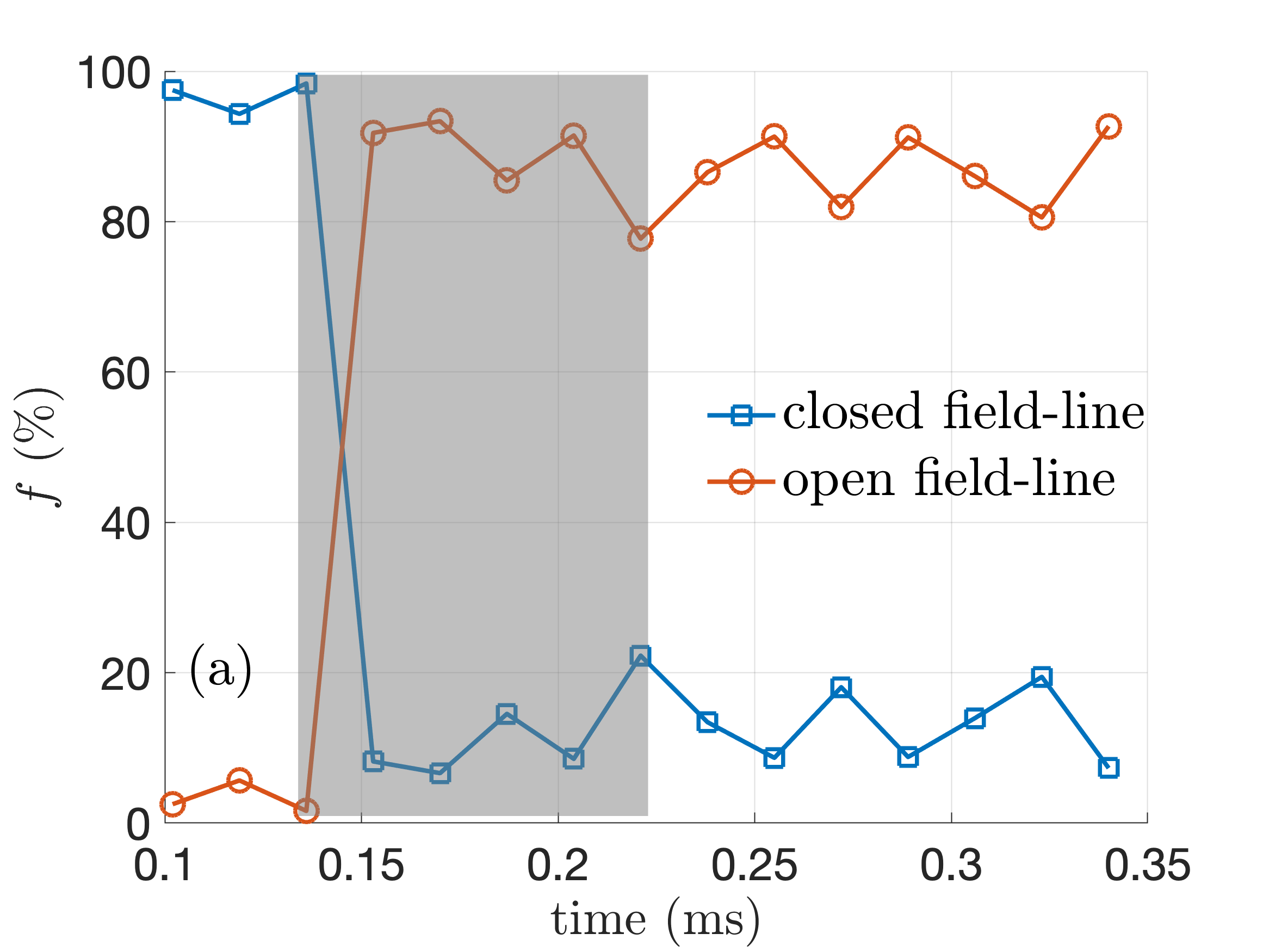}
    \includegraphics[width=0.45\linewidth]{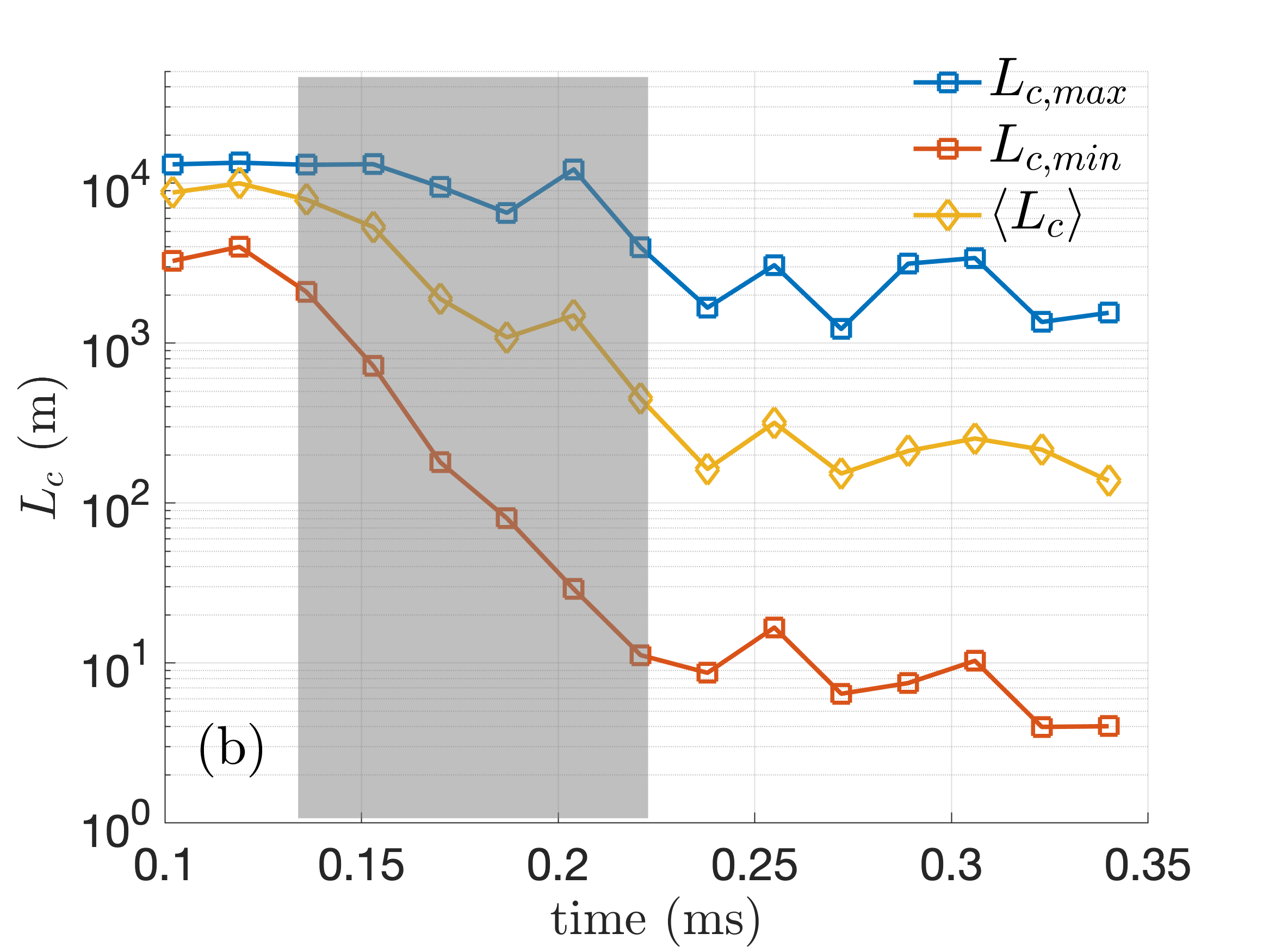}
    \caption{Temporal evolution of (a) open vs closed field-lines on $\psi_n=0.95$ surface, and (b) the maximum, minimum and flux surface averaged connection length of open field-lines.}
    \label{fig:fieldline_q95}
\end{figure}

The change of magnetic topology in our thermal quench simulation can also be verified by examining the statistics of field-lines on an initially closed magnetic flux surface.
To get a better statistics, a total of 3584 spatial points uniformly sampled along the field-line and toroidal directions on $\psi_n=0.95$ surface are selected as the starting points. The percentages of closed versus open field-lines, as well as the maximum, minimum, and averaged magnetic connection length $L_c$ of open field-lines at different times of the simulation are 
illustrated in Figure~\ref{fig:fieldline_q95}.
Prior to $t=0.136$~ms, the majority ($\geq 95\%$) of the field-lines on $\psi_n=0.95$ surfaces are ``closed", indicating that the pedestal top plasma is well-confined magnetically as there are only a few percents of the total field-lines leaking into the SOL region due to turbulence induced magnetic perturbation. Shortly after the thermal quench onset, the ``closed" field-line population drops to about $10\%$ and stays at the similar level for the rest of the simulation, suggesting that the $\psi_n=0.95$ surface is quickly distorted and the pedestal top plasma is likely directly connect to the divertor target plates and/or wall (outer boundary) during the thermal quench. The rapid drop of $L_{c,max},L_{c,min}$ and $\langle L_c\rangle$ after $t=0.136$~ms implies the magnetic fluctuation level increases substantially; and the trending continues even after the outflow from the core stops at $t=0.221$~ms.



\section{Radial particle and heat flux analysis}\label{sec:flux}

In this section, we will quantitatively analyze radial transport contributions from $E\times B$ turbulent convection process and parallel advection/conduction process (i.e., magnetic flutter effect). 
The radial particle flux $\Gamma_{n,r}$ and heat flux $Q_{p_\alpha,r}$ in BOUT++ turbulence simulations with the presence of perturbed magnetic fields is
\begin{equation}
    \Gamma_{n,r}=n\frac{(\vec{b}_0\times\nabla\phi)_r}{B_0}+nV_{\parallel i}\frac{\left(\vec{b}_0\times \nabla A_\parallel\right)_r}{B_0}.
\end{equation}
Here the first term on the right-hand-side is due to the radial component of $E\times B$ convection (i.e., cross-field turbulent transport) while the second term is caused by perturbed magnetic field along radial direction (i.e., parallel advection process). Analogously, for radial heat flux $Q_{p_\alpha,r}$
\begin{equation}\label{eq:qpr}
    Q_{p_\alpha,r}= \frac{3}{2}p_\alpha \frac{(\vec{b}_0\times\nabla\phi)_r}{B_0}+\frac{5}{2}p_\alpha V_{\parallel \alpha}\frac{\left(\vec{b}_0\times \nabla A_\parallel\right)_r}{B_0}+\kappa_{\parallel \alpha}\frac{\left(\vec{b}_0\times \nabla A_\parallel\right)_r}{B_0}\cdot \nabla T_\alpha,
\end{equation}
where the third term on the right-hand-side is the radial projection of parallel conductive heat flux in the presence of perturbed magnetic field. Note that in our model, the classical perpendicular heat flux due to collision is neglected as $\kappa_\perp\ll\kappa_\wedge\ll\kappa_\parallel$ for both ions and electrons.



\begin{figure}[h]
    \centering
    \includegraphics[width=0.45\linewidth]{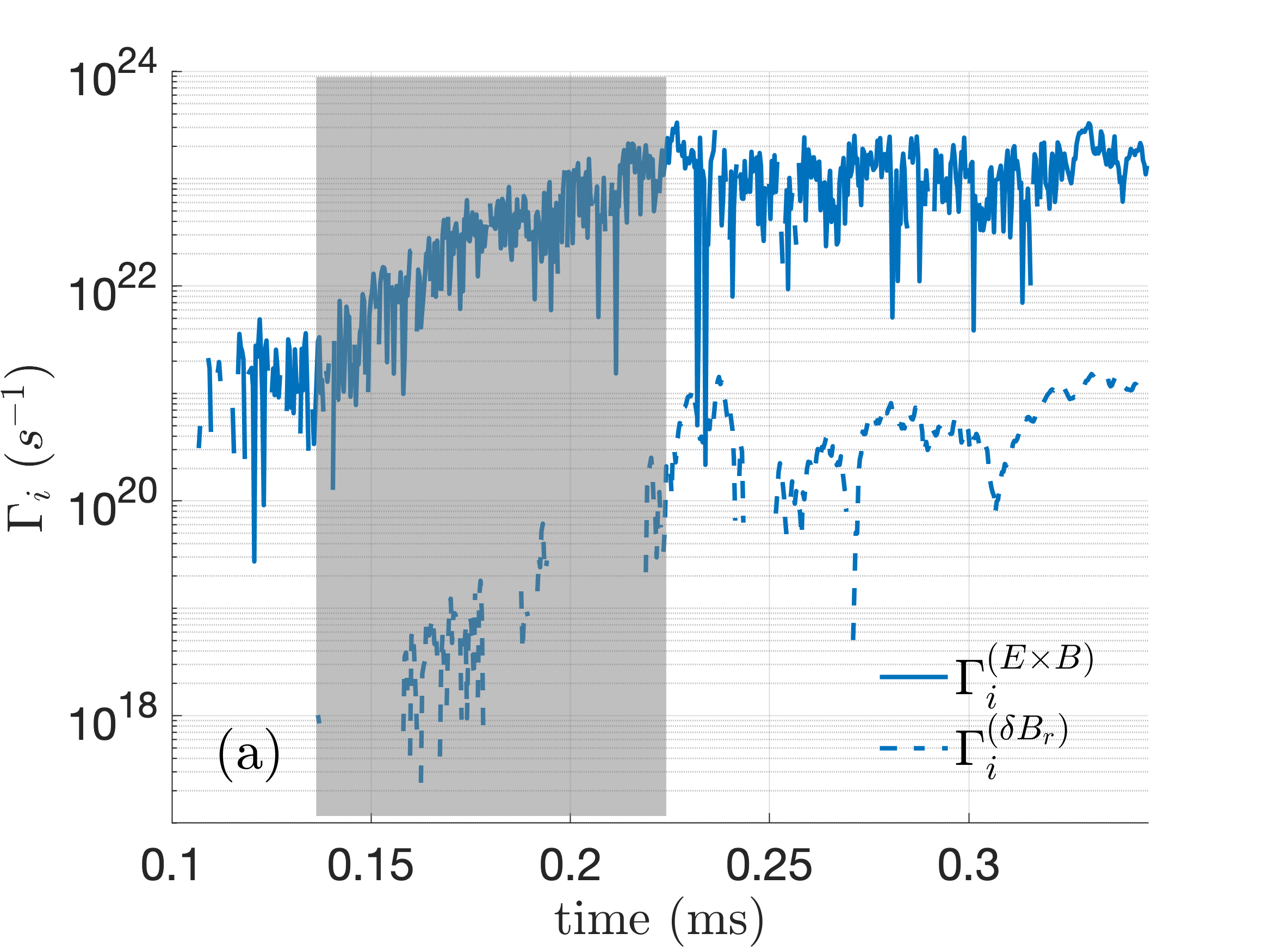}
    \includegraphics[width=0.45\linewidth]{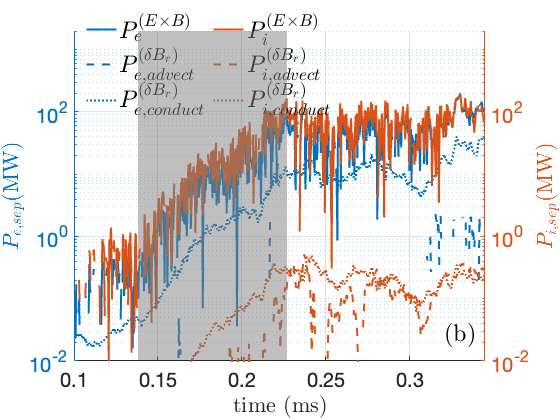}
    \caption{Temporal evolution of (a) particle, and (b) power across the separatrix.}
    \label{fig:flux_sep}
\end{figure}

We first evaluate the total radial flux across the separatrix since only the flux across the separatrix ends at the divertors and wall.
Figure~\ref{fig:flux_sep} shows the total particle and power across the separatrix with individual contributions from $E\times B$ convection and magnetic flutter effect. Not surprisingly, the perturbed magnetic field has minimum impact ($<0.1\%$) on the particle transport. This is because (1) the quasi-neutral condition forces electrons to move along with ions in our model and (2) ion parallel advection (i.e., $nV_{\parallel i}$) is weak inside the separatrix as we assume the equilibrium current $J_{\parallel 0}$ is all carried by electrons (i.e., $V_{\parallel i 0}=0$).
Unlike the $E\times B$ convection process which treats electrons and ions equally, the magnetic flutter effect often weights differently for electron and ion thermal transport. The parallel advection heat flux contributions (i.e., second term on the RHS of Equation~\ref{eq:qpr}) are negligible, again, as a consequence of vanishing equilibrium current at the separatrix assumption in our model (i.e., $V_{\parallel e 0}= V_{\parallel i 0}=0$ at $\psi_n=1$).
For the parallel conductive flux, because electrons are much more mobile (or lighter) than ions, electron parallel conductive heat flux is roughly two orders of magnitude larger than ion parallel conductive heat flux (e.g.,$\kappa_{\parallel e}/\kappa_{\parallel i}\sim (m_i/m_e)^{1/2}\sim 57$ for deuterium plasma). Therefore, in a perturbed magnetic field, electrons often have a much larger radial heat flux component from the magnetic flutter effect than ions.
In this study, the turbulent $E\times B$ convection dominates the radial heat transport. This is particularly true for ions as magnetic flutter only contribute less than $1\%$ of total ion radial heat flux for the entire simulations. However, for electrons, magnetic flutter effect does play an important role. It accounts for roughly $10\%$ of the total electron radial heat flux prior to the thermal quench onset (i.e., $t=0.136$~ms) and this number gradually increases to $30\%$ -- only a factor of 2 smaller than $E\times B$ convection process, in the later stage as the magnetic field becomes fully stochastic. This finding is in a good agreement with BOUT++'s previous study on DIII-D type-I ELM that magnetic flutter effect can substantially enhance the radial heat transport and as a consequence, the total energy loss increases by $\sim33\%$~\cite{xia2015nonlinear}. It is also consistent with the recent nonlinear extended MHD study on ELM dynamics~\cite{cathey2021comparing} and the electromagnetic gyrokinetic scrape-off-layer turbulence study~\cite{mandell2022turbulent}. Therefore, it is possible that under certain plasma conditions, electron radial thermal transport may be dominated by the stochastic field transport instead of turbulent $E\times B$ process during the thermal quench phase.


\section{Conclusion}\label{sec:con}
In this paper we report the thermal quench simulation performed with BOUT++ six-field electromagnetic turbulence model with a particular focus on the governing physics of edge plasma transport as well as the divertor heat load in this transient event.
The simulation is carried out for a generic, quiescent DIII-D lower-single-null H-mode plasma with a short period of intense particle and energy injection at the pedestal top to mimic the plasma outflow from the core region when thermal quench is triggered.
The nonlinear simulation result quantitatively reproduces several important edge plasma features observed in experiments. For example, the divertor heat-load surging and heat-flux width broadening. \revision{Experimental observation confirms an order of magnitude heat load increasing~\cite{hender2007mhd,lehnen2013disruption} and a few times divertor heat flux width broadening~\cite{lehnen2015disruptions} during the thermal quench.} While in our study, the maximum heat load on the outer divertor target plate increases 50 times and the heat flux width expands 4 times. \revision{Similar to the experiments, simulation also shows that the divertor heat load rise time $\tau_r$ roughly matches the thermal quench duration $\tau$. Our analysis indicates that this is a result from divertor heat load is governed by electron parallel conduction process at the early stage.}
More interestingly, the divertor heat footprint alters from quasi-coherent pattern to striation pattern after thermal quench occurs. 

The dramatic increasing of divertor heat load and broadening of heat flux width are tied to the enhanced turbulence activities. In the simulation, the maximum turbulent fluctuation level increases approximately 6 times, and the turbulent region expands from near the separartix to the entire simulation domain. The turbulence characteristics are also changing. As the dominant modes shift to lower wave-vector $k$, larger eddies and filaments provide a more effective radial transport channel. The temporal evolution of heat load on the outer divertor (downstream) is found to be strongly correlated with the turbulence activity at outer board mid-plane (upstream). By estimating the characteristic transport times and comparing the values to the observed lagging time, we conclude that electron parallel thermal conduction is the dominant divertor heat exhaust mechanism.

The enhanced plasma turbulence also comes with the amplified magnetic perturbation that causes the late-appeared striation heat load pattern on the divertor. Our field-line tracing analysis indeed shows that the intact magnetic surfaces starts to deform and break along with the thermal quench process from edge to core, and eventually the magnetic field becomes fully stochastic when $|\delta B_r/B_0|\sim 10^{-4}$. Therefore, shortly after thermal quench onsets, the destroyed magnetic flux surfaces are no longer able to confine plasma in the pedestal region; instead, core plasma now can directly connect to the plasma facing component so that the parallel transport combined with finite perturbed $B_r$ (i.e., magnetic flutter effect) may overwhelm the other radial transport mechanisms. Our further analysis suggests that in our simulation, the turbulent $E\times B$ radial transport dominates the magnetic flutter effect  in terms of transport particles and energy across the separatrix. However, the magnetic flutter effect does facilitate the electron radial heat transport as it contributes roughly $30\%$ of the total energy carried by electrons across the separatrix in the late stage. 

\revision{The authors would like to remark a few caveats of this study. For instance, the bootstrap current is assumed stationary, and the flux-limited parallel heat flux model is used in current study. Giving the substantial change of plasma profiles, the former assumption is questionable. Also, the additional coefficient $\alpha$ in flux-limited expression could impact the relative roles of the turbulent $E\times B$ transport and the magnetic flutter effect. Future thermal quench simulations are planned to include a self-consistent bootstrap current model, to employ a better Landau-fluid/kinetic~\cite{wang2019landau} parallel heat flux closure, and to extend the simulation domain to full torus. These improvements will highlight the peeling drive of the edge instability, which may result in a lower-$n$ mode dominated turbulence with even larger magnetic fluctuations such that the electron radial heat transport is primarily carried by the stochastic conduction process.}



\section*{Acknowledgments}
We thank the U.S. Department of Energy Office of Fusion Energy
Sciences and Office of Advanced Scientific Computing Research for
support under the Tokamak Disruption Simulation (TDS) Scientific
Discovery through Advanced Computing (SciDAC) project, both at
Lawrence Livermore National Laboratory (LLNL) under Contract
DE-AC52-07NA27344 and at Los Alamos National Laboratory (LANL) under
contract No. 89233218CNA000001. This research used resources of the
National Energy Research Scientific Computing Center, a DOE Office of
Science User Facility supported by the Office of Science of the
U.S. Department of Energy under Contract No. DE-AC02-05CH11231.
LLNL-JRNL-845128


\section*{Reference}
\bibliography{thermal_quench}

\end{document}